\documentclass[twocolumn,prc,showpacs,preprintnumbers,amsmath,amssymb,superscriptaddress]{revtex4}

\usepackage{graphicx}
\usepackage{dcolumn}
\usepackage{bm}

\begin{document}

\title{Formation of light exotic nuclei in low-energy multinucleon transfer reactions}

\author{V.I.~Zagrebaev}
\affiliation{Flerov Laboratory of Nuclear Reactions, JINR, Dubna, Moscow Region, Russia}
\author{B.~Fornal}
\affiliation{The Niewodniczanski Institute of Nuclear Physics, Polish Academy of Sciences, Krakow, Poland}
\author{S.~Leoni}
\affiliation{Dipartimento di Fisica, University of Milano, Milano, Italy}
\author{Walter~Greiner}
\affiliation{Frankfurt Institute for Advanced Studies, J.W.~Goethe-Universit\"{a}t, Frankfurt, Germany}

\date{\today}

\begin{abstract}
Low-energy multinucleon transfer reactions are shown to be very effective tool for the production and spectroscopic study of light exotic nuclei. The corresponding cross sections are found to be significantly larger as compared with high energy fragmentation reactions. Several optimal reactions for the production of extremely neutron rich isotopes of elements with $Z=6\div 14$ are proposed.
\end{abstract}
 \maketitle

\section{Motivation}\label{In}

Multinucleon transfer reactions occurring in low-energy collisions of heavy ions are currently considered as the most promising method for the production of new heavy (and superheavy) neutron-rich nuclei, unobtainable by other reaction mechanisms. These reactions can be used both for the production of new neutron-rich isotopes of transfermium elements (where only proton-rich nuclei located on the left side from the stability line have been synthesized so far) and new neutron-rich nuclei located along the closed neutron shell N=126 \cite{Zag08prl} (area of the nuclear map having the largest impact on the r process of astrophysical nucleosynthesis). Cross sections of these reactions are predicted to be rather large, making it possible to perform the corresponding experiments at available accelerators. The only problem here is the separation of heavy transfer reaction fragments, although proper separators are being designed and manufactured now in several laboratories.

\begin{figure}[ht]
\begin{center}\resizebox*{5.0 cm}{!} {\includegraphics{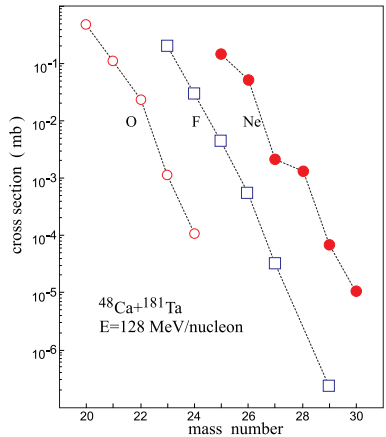}}\end{center}
\caption{(Color online) Cross sections for the production of oxygen, fluorine and neon isotopes in fragmentation of
$^{48}$Ca with $E=128$~MeV/nucleon on $^{181}$Ta target \cite{Kwan06}).
\label{fragm}}
\end{figure}

On the contrary, fission reactions and high energy fragmentation processes are successfully used for the production of neutron-rich medium mass and light exotic nuclei, correspondingly. Great progress here was done lately and dozens of new nuclei have been discovered, mainly at the laboratories of NSCL MSU \cite{Michigan}, RIKEN \cite{RIKEN} and  GSI \cite{GSI}.

The disadvantage in producing light exotic nuclei in fragmentation reactions relies mainly on the fact that in this case one uses beams of relatively heavy species, which are rather expensive if one wants to produce them with high intensity. Secondly, the cross section for production of exotic nuclei in fragmentation processes drops down very fast when moving away from the stability line, as it is shown, for example, in Fig.  \ref{fragm}.

One of the main objectives in the production of exotic nuclei is their spectroscopic study.
In particular, gamma spectroscopic studies exploiting deep-inelastic heavy-ion reactions look quite promising \cite{Bro06}.
Such reactions have been used successfully to study the yrast structure of hard-to-reach, neutron-rich nuclei in the vicinity of $^{36}$S \cite{For94,Lia02}, $^{48}$Ca \cite{Jan02,Mon11}, $^{64}$Ni \cite{Bro12, Rec12}, $^{76}$Ge \cite{Toh13}, $^{82}$Se \cite{Jon07}, $^{124}$Sn \cite{Bro92}, $^{208}$Pb \cite{For01,Cie12} and $^{232}$Th \cite{Coc99}. It was done by employing thick-target gamma-gamma coincidence technique with large germanium detector arrays: in such cases the resolving power of the arrays has proven sufficient to extract detailed information from coincidence data sets with large statistics, even for weak reaction channels. Alternatively, thin-target gamma-reaction product coincidence method was used, with gamma array coupled to magnetic spectrometer which provides full isotopic identification of reaction fragments, e.g., CLARA/AGATA+PRISMA \cite{Gad04,Gad11, Mon11a,Val09,Lou13}, EXOGAM+VAMOS \cite{Rej11,Bha08,Nav14}.

In view of that, one might expect that low-energy multinucleon-transfer reactions  may also serve as a tool for the production and investigation of very light exotic nuclei, a method which has not been applied so far. The idea would be to use a light and neutron-rich beam on a heavy target. The combination of a large acceptance magnetic spectrometer with a high efficiency and high resolution multidetector array for $\gamma$ spectroscopy would be a key instrument in such study.

Unfortunately, there is almost no (or very fragmentary) experimental information on the production cross sections of light reaction fragments formed in multinucleon transfer processes induced by light ions  on medium mass or heavy targets.
Also, there is no appropriate theoretical model (adjusted for description of such reactions) which could be used for accurate predictions of these cross sections. The well known GRAZING code \cite{GRAZING} describes properly only few neutron transfer channels, but it strongly underestimates the channels with proton transfers (see below).

In this paper we use the model based on the Langevin type equation of motions \cite{ZG05,ZG08} for the description of multinucleon transfer reactions with light heavy ions ($A\sim 20$) and for the prediction of the corresponding cross sections. This model has been developed originally for analysis of deep inelastic scattering and fusion-fission reactions occurring in collisions between heavy ions and it describes well these processes. However, it has never been applied for the description of collisions induced by light ions. Therefore, as the first step, we analyzed within the presented model available (not numerous) experimental data on light ion collisions and we showed that the model works reasonably well. Then, we considered low-energy multinucleon transfer reactions for several light ions as projectiles scattered on uranium target. The calculated cross sections for the production of light exotic nuclei in these reactions have been compared with those observed in high energy fragmentation processes.

\section{The Model}\label{Model}

Description of mass transfer in damped collisions of heavy ions is a rather difficult theoretical problem not solved yet completely.
Several simplified models for qualitative description of such reactions have been proposed in the past, namely, the
Focker-Planck \cite{Nor74} and master equations \cite{Moretto75} for the corresponding distribution function and the Langevin
equations \cite{Frobrich88}. Later more sophisticated semiclassical approaches \cite{Vigezzi89,Zag90,Winther95} have been also proposed.
The well known GRAZING code \cite{GRAZING} for description of nucleon transfer reactions in heavy ion collisions is available on the market
(now it is possible to run this code directly at the NRV web-site \cite{NRV_grazing}). The semiclassical model used by this code
describes quite well few nucleon transfer reactions (see, for example, review paper \cite{Corradi09}).
However, the multinucleon transfers are not reproduced within this model, it gives too narrow mass distributions of reaction fragments
because the damped reaction channels with large kinetic energy loss are not included in the model.
Recently the first successful attempt was done of using the microscopic approach based on the TDHF theory for numerical analysis
of multinucleon transfer reactions in low-energy collisions of heavy ions \cite{Yabana13}.
This approach (in spite of time consuming calculations) looks very promising.

Here we use the model based on the coupled Langevin-type dynamical equations of motion \cite{ZG05,ZG08} proposed for simultaneous description
of multinucleon transfer, quasi-fission and fusion-fission reaction channels (difficult-to-distinguish experimentally in many cases).
The adiabatic multi-dimensional potential energy surface calculated within the extended version \cite{extTCSM} of the two center shell model
is a key object of this approach which regulates the whole dynamics of low-energy nucleus-nucleus collision.
Calculations performed within the microscopic time-dependent Schr\"{o}dinger equations \cite{ZSG07} have clearly demonstrated
that at low collision energies of heavy ions nucleons do not ``suddenly jump'' from one nucleus to another. Instead of that, the wave functions
of valence nucleons occupy the two-center molecular states spreading gradually over volumes of both nuclei.
The same adiabatic dynamics of low-energy collisions of heavy ions was found also within the TDHF calculations \cite{Umar08,Simenel12, Yabana13}.
This means that the perturbation models based on a calculation of the sudden overlapping of single-particle wave functions of transferred nucleons
(in donor and acceptor nuclei, respectively) cannot be used for description of multinucleon transfers in low-energy heavy-ion damped collisions.

The distance between the nuclear centers $R$ (corresponding to the elongation of a mono-nucleus when it is formed),
dynamic spheroidal-type surface deformations $\delta_1$ and $\delta_2$, the neutron and proton asymmetries,
$\eta_N = (2N-N_{CN})/N_{CN}$, $\eta_Z = (2Z-Z_{CN})/Z_{CN}$ (where $N$ and $Z$ are the neutron and proton numbers in one of the fragments,
whereas $N_{CN}$ and $Z_{CN}$ refer to the whole nuclear system) are the most relevant degrees of freedom for the description of
mass and charge transfers in low-energy collisions of heavy ions.
For all the variables, with the exception of the neutron and proton transfers, we use the usual Langevin equations of motion
with the inertia parameters, $\mu_R$ and $\mu_{\delta}$, calculated within the Werner-Wheeler approach \cite{Davies76}
\begin{equation}
\frac{dq_i}{dt}=\frac{p_i}{\mu_i},\,\,\frac{dp_i}{dt}=\frac{\partial V_{\rm eff}}{\partial q_i}-\gamma_i\frac{p_i}{\mu_i}+\sqrt{\gamma_i T}\Gamma_i(t).\label{Langevin}
\end{equation}
Here $q_i$ is one of the collective variables, $p_i$ is the corresponding conjugate momentum, multi-dimensional potential energy $V_{\rm eff}$
includes the centrifugal potential,
$T = \sqrt{E^*/a}$ is the local nuclear temperature, $E^* = E_{\rm c.m.}-V_{\rm eff}(q_i;t)-E_{\rm kin}$ is the excitation energy,
$\gamma_i$ are the appropriate friction coefficients, and $\Gamma_i(t)$ are the normalized random variables with Gaussian distribution.
The quantities $\gamma_i$, $E^*$ and $T$ depend on all the coordinates and, thus, on time
(evidently all them are equal to zero at approaching reaction stage at large values of $R$).

Nucleon exchange (nucleon rearrangement) can be described by the inertialess Langevin type equations of motion derived from the master equations
for the corresponding distribution functions \cite{ZG05,ZG08}
\begin{eqnarray}\label{etanz}
&\displaystyle\frac{d\eta_N}{dt} = \frac{2}{N_{CN}} D^{(1)}_N+\frac{2}{N_{CN}}\sqrt{D^{(2)}_N}\Gamma_N(t),\\
&\displaystyle\frac{d\eta_Z}{dt} = \frac{2}{Z_{CN}} D^{(1)}_Z+\frac{2}{Z_{CN}}\sqrt{D^{(2)}_Z}\Gamma_Z(t).\nonumber
\end{eqnarray}
Here $D^{(1)}$, $D^{(2)}$ are the transport coefficients. We assume that sequential nucleon transfers play a main role in
mass rearrangement. In this case
\begin{eqnarray}\label{dd}
D^{(1)}_{N,Z} &=& \lambda_{N,Z}^{(+)}(A\to A+1)- \lambda_{N,Z}^{(-)}(A\to A-1), \\
D^{(2)}_{N,Z} &=& \frac{1}{2}[\lambda_{N,Z}^{(+)}(A\to A+1)+\lambda_{N,Z}^{(-)}(A\to A-1)],\nonumber
\end{eqnarray}
where the macroscopic transition probabilities $\lambda_{N,Z}^{(\pm)} (A\to A^\prime =A\pm 1)$ depend on the
nuclear level density \cite{Nor74,Moretto75}, $\lambda_{N,Z}^{(\pm)} = \lambda_{N,Z}^0\sqrt{\rho(A\pm 1) /
\rho(A)}$ and $\lambda_{N,Z}^0$ are the neutron and proton transfer rates.
The nuclear level density $\rho\sim exp(2\sqrt{aE^*})$ depends on the excitation energy $E^*$ and, thus, the transition probabilities,
$\lambda_{N,Z}^{(\pm)}$, are also coordinate and time dependent functions.

The first terms on the r.h.s. of Eqs.(\ref{etanz}), $D^{(1)}_N\sim \partial V/ \partial N$ and $D^{(1)}_Z\sim \partial V/ \partial Z$,
drive the system to the configuration with minimal potential energy in the $(Z,N)$ space (see below Fig.\ \ref{Ne_Mo}),
i.e., to the optimal Q-value of nucleon rearrangement.
The second terms in these equations, $\sim D^{(2)}_{N,Z}$, describe a diffusion of neutrons and protons in the configuration of two overlapped nuclei.

For separated nuclei the nucleon exchange is still possible (though it is less probable) and has to be taken into account in Eqs.~(\ref{etanz}).
We use the following final formula for the transition probabilities
\begin{equation}
\lambda_{N,Z}^{(\pm)}=\lambda^0_{N,Z}\sqrt{\frac{\rho(A\pm 1)}
{\rho(A)}}P_{N,Z}^{tr}(R,A\to A\pm 1).\label{transfer}
\end{equation}
Here $P_{N,Z}^{tr}(R,A\to A\pm 1)$ is the probability of one nucleon transfer (neutron or proton), which depends on the distance between the
nuclear surfaces and the nucleon separation energy. This probability goes exponentially to zero at $R\to\infty$ and it is
equal to unity for overlapping nuclei. The simple semiclassical formula is used for the calculation of $P_{N,Z}^{tr}$ (see \cite{ZG05,ZG08}).
Thus, Eqs.~(\ref{etanz}) -- (\ref{transfer}) define a continuous change of charge and mass asymmetries during the whole
process of nucleus-nucleus collision (obviously, $d\eta_{N,Z} / dt \to 0$ for far separated nuclei).

In our approach we distinguish the neutron and proton transfers (it is important for prediction of the yields of different isotopes of a given element).
At the approaching stage (for separated nuclei) the probabilities for neutron and proton transfers are different.
The Coulomb barrier for protons leads to faster decrease of their bound state wave functions outside the nuclei,
and, in general, $P_{Z}^{tr}(R>R_1+R_2,A\to A\pm 1) < P_{N}^{tr}(R>R_1+R_2,A\to A\pm 1)$.
However, for well overlapped nuclei single particle motions of protons and neutrons are rather similar,
and we assume that the neutron and proton transfer rates are equal to each other, i.e.,
$\lambda_N^0 = \lambda_Z^0 = \lambda^0/2$, and both are the parameters of the model (i.e., they are not derived from some microscopic calculations).
The model describes quite properly \cite{ZG08} experimental difference in the cross sections of pure neutron and proton transfers
in the case of heavy ion collisions \cite{Corradi09}.

The nucleon transfer rate, $\lambda^0$, is the fundamental quantity of low-energy nuclear dynamics.
However its value is not yet well determined.
For the first time the value of $\lambda^0$ was estimated roughly in Refs.\cite{Nor74,Moretto75} to be about $10^{22}$~s$^{-1}$.
In our previous studies we found that the value of the nucleon transfer rate of about $(0.05 - 0.1)\cdot 10^{22}$~s$^{-1}$ is quite sufficient
to reproduce experimental data on the mass distributions of reaction products in several heavy-ion damped collisions \cite{ZG05,ZG08}.
However this quantity is still rather uncertain. Its energy (and temperature) dependence was not studied yet.
Also it is not clear how it depends on masses of colliding nuclei.
More experimental data on multinucleon transfer reactions at different collision energies and for different colliding ions are needed
to determine the nucleon transfer rate accurately.
For all the reactions analyzed below we fixed the value of $\lambda^0 = 0.05\cdot 10^{22}$~s$^{-1}$.
Note that the larger is the value of $\lambda^0$ the wider are the mass and charge distributions of reaction fragments.

Another uncertain quantity of low-energy nuclear dynamics is the nuclear friction (nuclear viscosity) responsible for the kinetic energy loss
in heavy ion damped collisions. A great interest to these processes was shown 30 years ago. Those time, however, there was not appropriate theoretical
model for overall quantitative description of available experimental data on the mass, charge, energy and angular distributions of reactions products.
A number of different mechanisms have been suggested in the literature to be responsible for the energy loss in heavy ion collisions.
A discussion of the subject can be found, e.g., in \cite{Bass80,Treatise1,Abe96,Frob98}.
Microscopic analysis shows that nuclear viscosity may also depend strongly on nuclear temperature \cite{Hofmann}.
The uncertainty in the strength of nuclear viscosity (as well as its form-factor) is still large.

However all theoretical models as well as analysis of available experimental data conclude that the nuclear viscosity is rather large,
and it leads to the so-called overdamped collision dynamics of heavy ions. This means that for well overlapped nuclei kinetic energy stored
in all the degrees of freedom is rather low and excited nuclear system creeps slowly along the potential energy surface in the multi-dimensional
configuration space. As a result, the mass, energy and angular distributions of binary reaction products depend mainly on the form-factor
(e.g., on the radius) of friction forces, and not so much on the value of nuclear viscosity.
The strength parameter of nuclear friction as well as its form-factor are discussed in \cite{ZG05,ZG08}.

The double differential cross-sections of all the binary reaction channels are calculated as follows
\begin{equation}
\frac{d^2\sigma_{N,Z}}{d\Omega dE}(E,\theta) = \int_0^\infty{b
db}\frac{\Delta N_{N,Z} (b,E,\theta)}{N_{\rm tot}(b)}
\frac{1}{\sin(\theta)\Delta\theta\Delta E}. \label{cs}
\end{equation}
Here $\Delta N_{N,Z}(b,E,\theta)$ is the number of events (trajectories) at a given impact parameter $b$ in which a nucleus $(N,Z)$ is formed
in the exit channel with a kinetic energy in the region ($E,E+\Delta E$) and with a center-of-mass outgoing angle in the interval
($\theta,\theta+\Delta\theta$). $N_{\rm tot}(b)$ is the total number of simulated events for a given value of the impact parameter.
This number depends strongly on the low level of the cross section which one needs to reach in calculations. For predictions of rare events
with cross sections of 1~$\mu$b (primary fragments) one needs to test no fewer than $10^7$ collisions (as many as in a real experiment).

Expression (\ref{cs}) describes the mass, charge, energy and angular distributions of the {\it primary} fragments formed in the
binary reaction. Subsequent de-excitation cascades of these fragments via emission of light particles and gamma-rays in
competition with fission are taken into account explicitly for each event within the statistical model, leading to the {\it final}
distributions of the reaction products. The sharing of the excitation energy between the primary fragments is assumed here to be
proportional to their masses. This is also a debatable problem (see discussion below).

\section{Multinucleon Transfer Reactions In Collisions Of Light Nuclei (Analysis Of Available Experimental Data)}\label{experiment}

The model described above has not been used so far for analysis of low-energy collisions induced by relatively light heavy ions. It is then mandatory to  perform such analysis  before making any predictions within the model. However, as already mentioned, there are almost no experimental data on the isotopic yields (cross sections) of transfer reaction products (with identification of $Z$ and $A$) for collisions of low-energy light heavy ions ($A\sim 20$) with medium and heavy mass targets (high quality data of the desired precision were obtained only recently for medium mass projectiles and heavy targets \cite{Corradi09}).

\begin{figure}[ht]
\begin{center}\resizebox*{5.0 cm}{!} {\includegraphics{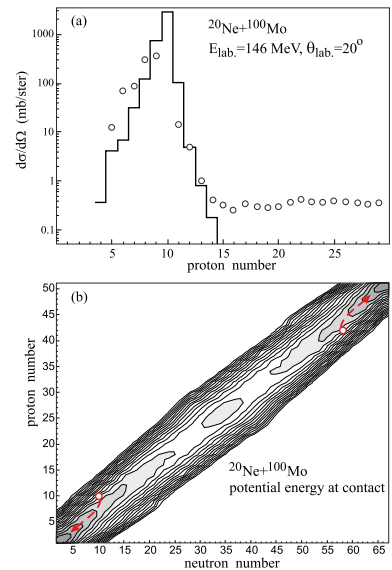}}\end{center}
\caption{(Color online) Upper panel: Charge distribution of reaction fragments in collisions of $^{20}$Ne with $^{100}$Mo
at $E_{\rm lab.}$=146 MeV. Experimental data (circles) are taken from \cite{Nagame84}. Bottom panel: Potential energy
of the nuclear system at contact configuration depending on nucleon rearrangement. Arrows indicate the most probable path
of nucleon transfers.
\label{Ne_Mo}}
\end{figure}

Of some help, however, is the work presented in Ref. \cite{Nagame84} where damped collisions of $^{20}$Ne with $^{100}$Mo have been studied at several energies slightly above the Coulomb barrier. Projectile-like fragments (PLF’s) were identified by their atomic numbers and the differential production cross sections were measured at angles near the grazing angle ($\theta_{\rm lab}^{\rm gr}\sim 30^{\rm o}$). Experimental charge distribution of reaction fragments is shown in Fig.\ \ref{Ne_Mo} along with the results of our calculations using the model described above.
The agreement is not perfect but not bad. Experimental charge distribution is very asymmetric in this reaction: stripping of protons from the projectile is more probable than their pick-up from the target. This behavior is reproduced quite well by the model and explained by the bottom panel of Fig.\ \ref{Ne_Mo}: potential energy of this nuclear system for its contact configuration (two touching nuclei) decreases just in the direction of nucleon transfer from lighter projectile to heavier target, thus increasing mass asymmetry in the exit channel. Such behavior (i.e., preferable evolution of nuclear system along valleys of driving potential) is generally inherent for damped collisions of heavy nuclei (e.g., well-known quasi-fission process), but, as we see, it can be attributed also to multinucleon-transfer processes in collisions induced by relatively light heavy ions on heavy targets.

\begin{figure}[ht]
\begin{center}\resizebox*{5.0 cm}{!} {\includegraphics{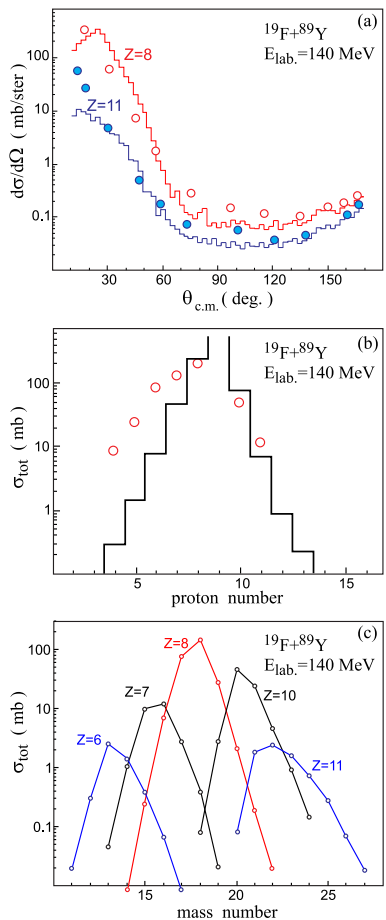}}\end{center}
\caption{(Color online) Angular (a), charge (b) and isotopic distributions (c) of projectile--like fragments
in collisions of $^{19}$F with $^{89}$Y at $E_{\rm lab.}$=140 MeV. Experimental data shown in (a) and (b) are taken from \cite{Mermaz86}.
\label{F_Y}}
\end{figure}

Very similar experiment on damped collisions of $^{19}$F  with $^{89}$Y was performed at 140 MeV beam energy \cite{Mermaz86}. Angular, energy and charge distributions of PLFs have been measured. Experimental data as well as the results of our calculations are shown in Fig.\ \ref{F_Y}. Agreement is about the same as for the previous reaction. Note that beside the dominating yields of PLFs at forward angles ($\theta_{\rm lab}^{\rm gr}\sim 24^{\rm o}$ for this reaction) there is a noticeable component with a wide (almost symmetric) angular distribution which is well reproduced by the model. These (rather rare) events of PLFs scattering to backward angles correspond to the trajectories with intermediate impact parameters $0<b<b_{\rm gr}$  when colliding nuclei are captured in the potential pocket and rotate but finally (owing to fluctuations) avoid fusion. On the bottom panel of Fig.\ \ref{F_Y} the calculated isotopic yields of PLFs are shown integrated over all angles. As can be seen, the cross sections for the production of light exotic nuclei in the considered reactions are rather high.

\begin{figure}[ht]
\begin{center}\resizebox*{6.5 cm}{!} {\includegraphics{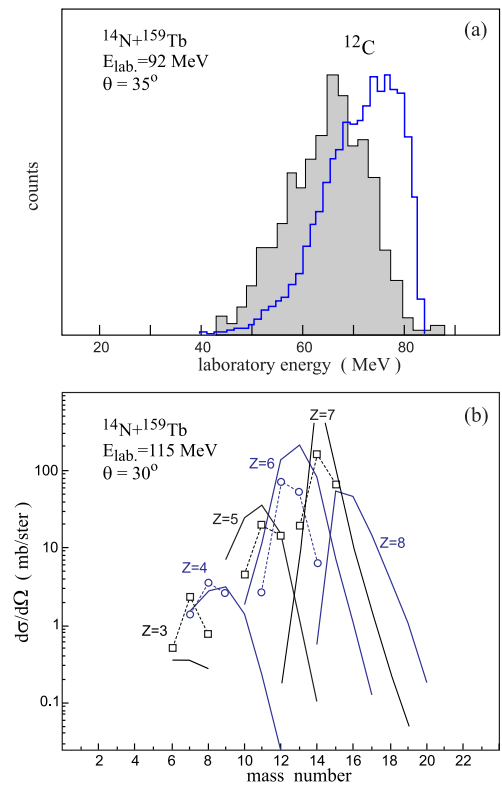}}\end{center}
\caption{(Color online) (a) Energy spectrum (linear scale) of $^{12}$C produced in the reaction $^{14}$N+$^{159}$Tb at $E_{\rm lab.}$=92 MeV.
Hatched area shows experimental data \cite{Balster1_87} and the histogram demonstrates result of calculation (the histograms are
equalized in vertical scale because an absolute normalization was not made in the experiment).
(b) Isotopic distributions of projectile-like elements formed in collisions of $^{14}$N with $^{159}$Tb at $E_{\rm lab.}$=115 MeV
and $\theta_{\rm lab.}=30^{\rm o}$. Solid lines show theoretical estimations. Experimental data (squares and circles connected
by dashed lines for $Z=3\div 7$) are taken from \cite{Balster87}.
\label{N_Tb}}
\end{figure}

In \cite{Balster1_87, Balster87}, a complete experimental study of the mechanisms of PLF production has been made for low-energy collisions of $^{14}$N with $^{159}$Tb. Coincident detection of K-Xrays of target-like fragments clearly demonstrates that at low collision energies the binary transfer reactions, which bring a dominant contribution to the yields of PLF heavier than lithium, dominate. The measured energy distributions of PLFs \cite{Balster1_87} demonstrate typical damped mechanism  of their formation with large dissipation of kinetic energy (see the upper panel of Fig.\ \ref{N_Tb}). This can be the reason that the model based on the Langevin-type equations of motion still describes quite satisfactory multinucleon transfer processes in reaction with so light projectiles. On the bottom panel of Fig.\ \ref{N_Tb} experimental \cite{Balster87} and theoretical differential cross sections are shown
for the production of PLFs in the reaction $^{14}$N+$^{159}$Tb at beam energy $E_{\rm lab.}$=115 MeV and $\theta_{\rm lab.}=30^{\rm o}$. Agreement between the results of theoretical calculations and experimental data is not so bad if one ignores the yields of very light fragments.

\section{Production of light exotic nuclei in low-energy collisions of heavy ions}\label{transtarget}

Keeping in mind that the model described in Section \ref{Model}  reproduces quite satisfactory the yields of projectile-like fragments formed in low-energy binary collisions of relatively light ions with medium mass and heavy targets, we tried to predict the cross sections for the production of light exotic nuclei in multinucleon transfer reactions and compare them with the corresponding high-energy fragmentation processes. We restricted our analysis by a search for optimal reactions  which produce  light neutron rich nuclei. It is absolutely clear that for this purpose one needs to test collisions of most neutron rich projectiles and targets. Even in such case there are too many combinations to be tested and we again restrict our analysis to the reactions on neutron-rich $^{238}$U target.

\begin{figure}[ht]
\begin{center}\resizebox*{4.9 cm}{!} {\includegraphics{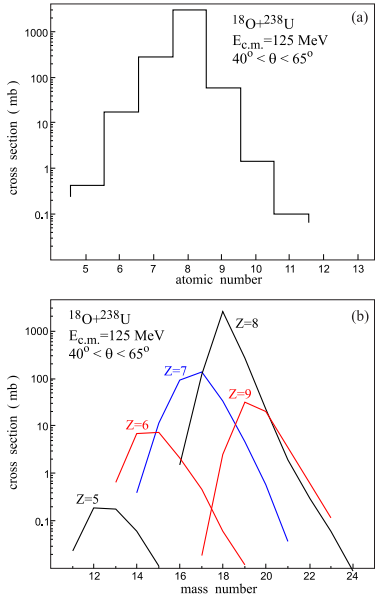}}\end{center}
\caption{(Color online) Charge (a) and isotopic (b) distributions of projectile-like fragments in collisions
of $^{18}$O with $^{238}$U at $E_{\rm c.m.}$=125 MeV. \label{O_U}}
\end{figure}
\begin{figure}[ht]
\begin{center}\resizebox*{4.9 cm}{!} {\includegraphics{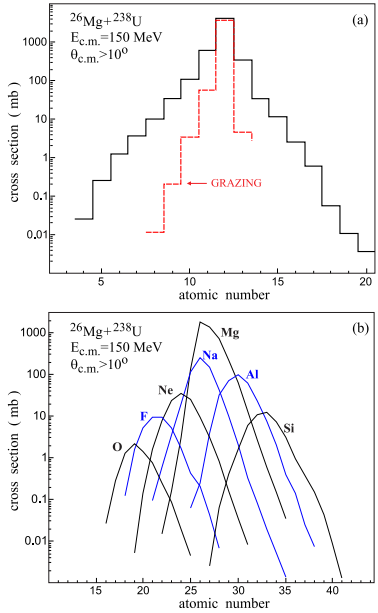}}\end{center}
\caption{(Color online) Charge (a) and isotopic (b) distributions of projectile--like fragments in collisions
of $^{26}$Mg with $^{238}$U at $E_{\rm c.m.}$=150 MeV. Dashed histogram in (a) shows the results of the GRAZING code.\label{Mg_U}}
\end{figure}
\begin{figure}[ht]
\begin{center}\resizebox*{5.0 cm}{!} {\includegraphics{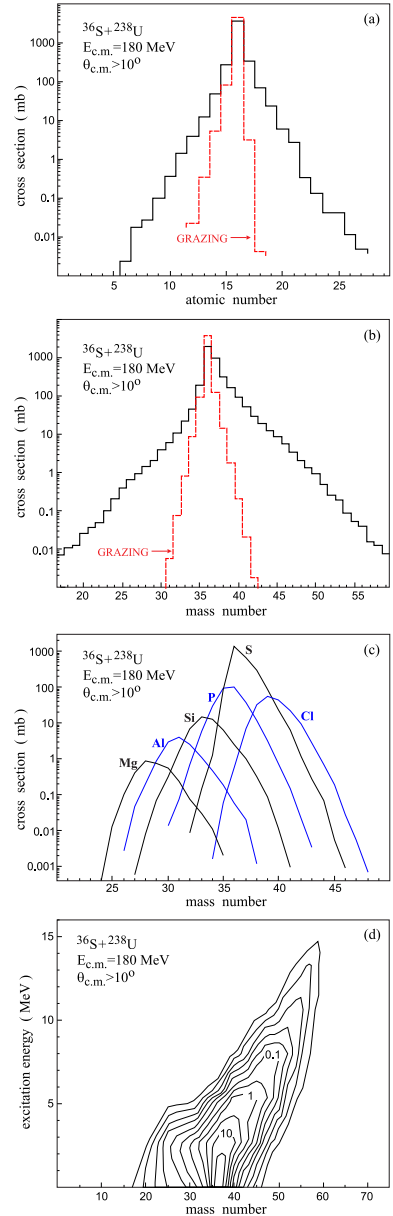}}\end{center}
\caption{(Color online) Charge (a), mass (b), isotopic (c) and excitation energy (d) distributions of projectile--like fragments in collisions of $^{36}$S with $^{238}$U at $E_{\rm c.m.}$=180 MeV. \label{S_U}}
\end{figure}

In Figs.\ \ref{O_U}, \ref{Mg_U} and \ref{S_U} the predicted cross sections are shown for the production of projectile-like fragments formed in multinucleon transfer processes of low-energy collisions induced by $^{18}$O, $^{26}$Mg and $^{36}$S projectiles on $^{238}$U  target. We compare our predictions with the similar calculations made by the GRAZING code \cite{GRAZING} which gives much narrower charge and mass distributions. However, it is well known that this code significantly underestimates the cross sections of proton transfers also for collisions of medium mass ions with heavy targets whereas the model used in our calculations describes such reactions reasonably well \cite{ZG05,ZG08}.

In all the figures the yields of primary PLF are demonstrated. Total excitation energy of PLFs and TLFs are rather high, even at low collision energies considered here. Usually, it is assumed that in heavy ion binary damped collisions the excitation energy is shared between the ejectiles proportionally to their masses (equal temperature of re-separated fragments in the exit channel). If it is true, then the light PLFs considered here should have rather low excitation energies (see Figs.\ \ref{S_U}) and, thus, no more than one neutron can be evaporated, shifting only negligibly the curves in Figs.\ \ref{O_U}, \ref{Mg_U} and \ref{S_U} toward lower masses.

The assumption about energy division has not been proven unambiguously by experiments. Extended discussion of the problem can be found, for example, in \cite{Toke92}. Note that the previous calculations of survival probabilities of excited primary PLFs and TLFs formed in collisions of heavier ions agree well with the corresponding experimental data if one assumes temperature equilibration in the exit channel \cite{ZG05,ZG08}.
However, keeping in mind this still unsolved problem of excitation energy sharing in very asymmetric combinations and a limitation of the statistical model for the description of decay probabilities of light excited nuclei, we restricted ourselves to the calculations of the cross sections for the production of primary PLFs shown in Figs.\ \ref{O_U}, \ref{Mg_U} and \ref{S_U}.

\begin{figure}[ht]
\begin{center}\resizebox*{7.5 cm}{!} {\includegraphics{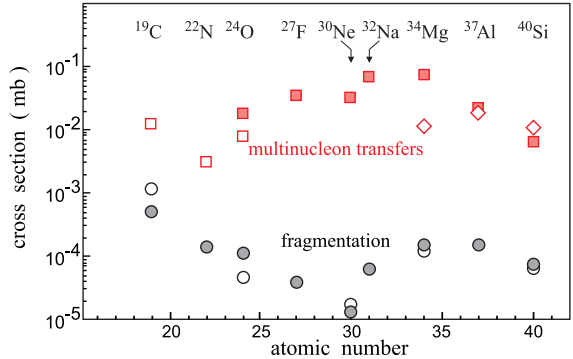}}\end{center}
\caption{(Color on line) Cross sections for the formation of light neutron rich nuclei in low-energy collisions of $^{18}$O (open rectangles),
$^{26}$Mg (filled rectangles) and $^{36}$S (open diamonds) with $^{238}$U target and in fragmentation of 128A MeV $^{48}$Ca
on $^{181}$Ta target \cite{Kwan06} (filled circles) and 345A MeV $^{48}$Ca on $^{9}$Be target \cite{RIKEN} (open circles).
 \label{tr_frag}}
\end{figure}

As can be seen from the obtained results, the cross sections for the formation of light neutron rich nuclei in low energy damped collisions of light heavy ions with heavy targets are significantly larger than the corresponding yields of these nuclei in high energy fragmentation processes.

In Fig.\ \ref{tr_frag} formation cross sections of several neutron rich nuclei (such as $^{19}$C, $^{24}$O, $^{30}$Ne, etc.) are compared for both the processes. The yields of these nuclei in the low-energy multinucleon transfer reactions are higher by about 2 orders of magnitude as compared with fragmentation reactions. Note that intensity of low-energy primary beams of such projectile as $^{18}$O, $^{26}$Mg and other can also be much higher than intensity of high energy beams used in the fragmentation reactions. Both factors make low-energy damped collisions of light heavy ions quite attractive for the production and study of light exotic nuclei just at presently available experimental facilities.

\section{Conclusion}\label{conclusion}

Within the model developed earlier for the description of damped collisions of heavy ions, we studied the multinucleon transfer reactions in low-energy collisions of light heavy ions with heavy targets. Comparison of theoretical calculations with (not numerous) available experimental data demonstrated rather good agreement.

Being inspired by this agreement, we calculated the cross sections for the formation of light exotic nuclei in low-energy collisions of $^{18}$O, $^{26}$Mg and $^{36}$S with $^{238}$U target.
The results of these calculations demonstrate that the yields of quite exotic light neutron-rich nuclei produced in the low-energy multinucleon transfer reactions are higher by about 2 orders of magnitude as compared with high energy fragmentation reactions. Thus the low energy damped collisions of light heavy ions with heavy targets look very promising and quite competitive to the fragmentation reactions for the production and study of light exotic nuclei.

\end{document}